\documentclass[prd,preprintnumbers,twocolumn,eqsecnum,floatfix,nofootinbib,a4paper,superscriptaddress]{revtex4-1}
\usepackage{color}
\usepackage{calc}
\usepackage{amsmath,amssymb,graphicx}
\usepackage{amssymb,amsmath}
\usepackage{tensor}
\usepackage{bm}
\usepackage{float}
\usepackage{microtype}
\usepackage{booktabs}
\usepackage{times}
\usepackage[varg]{txfonts}
\usepackage{subfigure}
\allowdisplaybreaks
\usepackage[utf8]{inputenc}
\usepackage{ulem}
\usepackage{svg}

\usepackage{comment}
\usepackage[nolist]{acronym}
\usepackage{tikz}
\usepackage{textcomp}
\usepackage[export]{adjustbox}
\usepackage{url}

\usetikzlibrary{arrows,shapes,trees,decorations.pathreplacing}
\usepackage[colorlinks,citecolor=blue]{hyperref}

\usepackage[utf8]{inputenc}
\usepackage{amsmath,amssymb}
\usepackage{comment}
\usepackage{amsmath}
\usepackage{braket}
\usepackage{mathrsfs}
\usepackage{mathtools}
\usepackage{array}
\usepackage{fancyhdr}
\usepackage{color}
\usepackage{bm}
\usepackage{url}

\usepackage{tikz}

\usepackage{aas_macros}

\allowdisplaybreaks

\newcommand{\codev}{%
   \overline{\nabla}
}

\begin{document}

\title{New Consistency Relations between Averages and Variances of \\Weakly Lensed Signals of Gravitational Waves}

\author{Morifumi Mizuno}
\affiliation{
Faculty of Science and Technology, Hirosaki University, 3 Bunkyo-cho, Hirosaki, Aomori 036-8561, Japan
}
\affiliation{
Department of Physics, University of Arizona, Tucson, Arizona 85721, USA
}
\author{Teruaki Suyama}
\affiliation{Department of Physics, Tokyo Institute of Technology, 2-12-1 Ookayama, Meguro-ku,
Tokyo 152-8551, Japan}
\author{Ryuichi Takahashi}
\affiliation{
Faculty of Science and Technology, Hirosaki University, 3 Bunkyo-cho, Hirosaki, Aomori 036-8561, Japan
}

\begin{abstract}
The lensing of gravitational waves (GWs) occurs when GWs experience local gravitational potential.
In the weak lensing regime, it has been reported that a simple consistency relation holds between the variances of the magnification and phase modulation.
In this paper, we present two additional consistency relations between the averages and variances of the weakly lensed GW signals in wave optics.
We demonstrate that these consistency relations are derived as the weak lensing limit of the full-order relations for the averages of the amplification factor and its absolute square.
These full-order relations appear to originate from energy conservation and the Shapiro time delay, and they are demonstrated to hold irrespective of the matter distribution.
\end{abstract}
%\keywords{Gravitational lensing: weak, Gravitational waves}
\maketitle
\section{Introduction}
The direct detection of gravitational waves (GWs) from binary black holes \citep{Abbott.Abbott.ea2016feb} and the detection of background GWs \citep{Agazie.Anumarlapudi.ea2023jun} has marked the onset of the GW astronomy era.
With ongoing and the expectation of future discoveries in the coming decade, our understanding of the Universe is set to reach new depths \citep{Bailes.Berger.ea2021may}.\text{\footnote{test}}

Gravitational lensing, which has been extensively studied in the context of light \citep{Bartelmann.Schneider2001jan,Mandelbaum2018}, also occurs in GWs \citep{Misner.Thorne.ea1973,Schneider.Ehlers.ea1992jan}.
Although the detection of lensed GW signals has not been reported to date, experimental efforts are underway to search for its evidence \citep{Abbott.Abe.ea2023apr}.
On the theoretical front, the lensing of GWs has been an active research subject. 
For example, gravitational lensing of GWs can enhance the amplitude of GWs, thereby causing the high tail for the redshifted mass distribution of black hole binaries \citep{Dai.Venumadhav.ea2017feb,Oguri2018nov}.
Note that there are distinct differences between the lensing of light and that of GWs, which is primarily due to the much longer wavelength of GWs. 
These differences give rise to the wave optics effect, primarily interference and diffraction, which can be used to extract complementary information about the lensing objects \citep{Ohanian1974jun,Nakamura1998feb,Takahashi.Nakamura2003oct,Caliskan.Ji.ea2023feb}. 
Specifically, lensing in wave optics is frequency dependent and involves a complex-valued quantity, i.e., the amplification factor, while in geometric optics, lensing effects arise simply due to light following the null geodesics in the curved spacetime.
Thus, measuring the amplification factor across a wide range of frequencies enables us to study the additional properties of lensing objects that cannot be captured in geometric optics.

In the weak lensing regime, the lensing of GWs is insensitive to structures smaller than the Fresnel scale \citep{Macquart2004aug,Takahashi2006jun}.
In other words, diffraction suppresses the lensing effect even if the lensing object is close to the line of sight as long as the object has a scale smaller than the Fresnel scale\footnote{When a point mass lens is considered, the length scale that must be compared with the Fresnel scale is the Einstein radius. This is equivalent to a comparison between the wavelength of GWs and the Schwarzshild radius of the lens.}. 
Since the Fresnel scale corresponding to typical GWs observed by ground-based detectors ($f\sim 1$Hz) is the order of 1 pc given that the distance between the GW source and observer is the cosmological distance scale, 
this feature can be exploited to probe the small-scale matter density fluctuations corresponding to the Fresnel scale of detectable GWs \citep{Takahashi2006jun,Oguri.Takahashi2020sep,Choi.Park.ea2021sep}.
If the observed GWs are enhanced due to strong lensing, the weak lensing signals superimposed on them would also be enhanced and more easily discerned \citep{Oguri.Takahashi2022aug}.
Weak lensing is based on the Born approximation and its precision is investigated by including the post-Born corrections \citep{Mizuno.Suyama2023aug}.
There, it is shown that the averages of the magnification and phase modulation become biased once the post-Born corrections are included.

In these weak lensing studies of GWs, it has been demonstrated that the variances of the magnification and phase modulation satisfy a universal and very simple relation \citep{Inamori.Suyama2021sep}.
While its physical meaning was not identified at the time, this relation provides a nontrivial connection between the real and imaginary parts of the amplification factor (thus, the consistency relation) and holds irrespective of the shape of the matter power spectrum.
In addition, another consistency relation for the real and imaginary parts of the amplification factor, i.e., the GW version of the Kramers-Kroning relation, has been reported \citep{Tanaka.Suyama2023aug}.

In this paper, we demonstrated the existence of two additional consistency relations for the averages and variances of the magnification and phase modulation. 
In doing so, we review the weak lensing of GWs in wave optics and show that the averages of the magnification and phase modulation are nonzero at the level of the post-Born approximation.
Then, we explain how the additional consistency relations hold and argue that these relations as well as the relation derived by \cite{Inamori.Suyama2021sep} can be understood as the weak lensing limit of more comprehensive relations that hold to infinite order in the gravitational potential. 
Importantly, one relation emerges as a consequence of the energy conservation law of GWs, and the second additional relation and a previously reported relation (Eq.~(\ref{eq: Inamori})) are attributed to the Shapiro time delay.
Interpreting lensing as a consequence of the Shapiro time delay appears to provide a physical explanation for the question raised by \cite{Inamori.Suyama2021sep}.

The rest of the paper is organized as follows. 
In section \ref{sec:weaklens}, the weak lensing of GWs is reviewed and the key quantities (i.e., the averages and variances of the magnification and phase modulation) are derived. 
In section \ref{sec: consistency}, the existence of two additional consistency relations is demonstrated and their physical meaning (energy conservation and the Shapiro time delay) as well as their significance in observations is discussed.
Section \ref{sec:cocl} concludes the paper.
Throughout this paper, we take $c=1$ and $\hbar=1$.

\section{Weak lensing of Gravitational waves}\label{sec:weaklens}
In most astronomical situations, perturbations to the relevant metric due to the presence of matter clumps are small, and the space-time metric is given as follows:
\begin{align}
    \label{eq:FRW metric}
    ds^{2}=&-\left(1+2\Phi\right)dt^{2}+\left(1-2\Phi\right)d{\bm x}^{2},
\end{align}
where $\Phi$ is the Newtonian gravitational potential.
In this case, the wave equation for the amplitude of GWs $\phi$ can be expressed as follows:
\begin{align}
    \label{eq: wave equation}
    \nabla^{2}\phi-(1-4\Phi)\frac{\partial^{2} \phi}{\partial t^{2}}=0,
\end{align}
where we assume that $\Phi$ varies very slowly with time and ignore its time derivative.
The derivation of this equation is predicated on certain assumptions, including the consideration of a small gravitational potential $|\Phi|\ll1$ and omission of polarization effects, as well as the assumption that the typical curvature radius induced by $\Phi$ is much larger than the wavelength of GWs. 
While the detail is beyond the scope of this paper, a rigorous derivation of the wave equation can be found in the literature \citep{Nakamura.Deguchi1999jan,Takahashi.Nakamura2003oct,Leung.Jow.ea2023apr}. 
Note that the expansion of the Universe is ignored in both Eqs.~(\ref{eq:FRW metric}) and (\ref{eq: wave equation}).
However, the inclusion of the expansion does not change these equations once $t$ and $\bm x$ are replaced with the conformal time and comoving distance with an associated redefinition of GWs due to attenuation of their amplitude as $\phi\to \phi/a$ \citep{Maggiore2018apr}. 

\begin{figure}[t]
    \centering
    %\includesvg[width=6cm,height=9cm,distort=false]{GWlensFig.svg}
    %\includesvg{GWlensFig.svg} 
    \includegraphics[width=8.5cm]{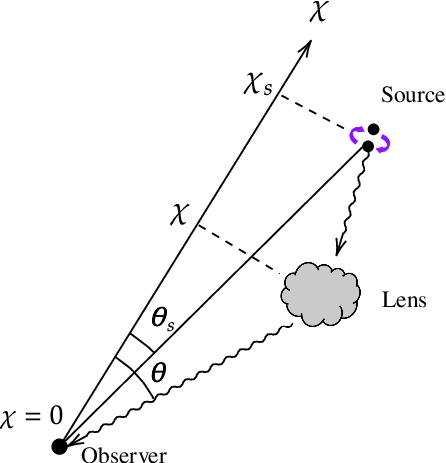} 
    \caption{ Gravitational lensing geometry. Following \citep{Nakamura.Deguchi1999jan}, we use the coordinate system in which the distance from the observer is $\chi$, and $\bm \theta$ is the two-dimensional vector perpendicular to the line of sight. The GW source is located at $(\chi_{s},\bm \theta_{s})$ where $|\bm \theta_{s}|\ll1$. 
    In the flat sky approximation, GWs reaching the observer are confined to the region $|\bm \theta|\ll1$.
    }
    \label{fig: GW lens}
\end{figure}

The lensing effect is commonly described in terms of the amplification factor, which is defined as the ratio of the lensed waveform to the unlensed waveform in the frequency domain, i.e., $F(\omega)\equiv\tilde{\phi}(\omega)/\tilde{\phi}_{0}(\omega)$ where the unlensed waveform is given by $\tilde{\phi}_{0}(\omega)=A e^{i\omega \chi}/\chi$, where $\chi$ is the distance from the GW source located at the origin.
Under the assumption that the typical wavelength of GWs is much smaller than the spatial variation of $F(\omega)$, Eq.~(\ref{eq: wave equation}) is rewritten as follows:
\begin{align}
    \label{eq: Amplification factor F}
    i \frac{\partial F}{\partial \chi}+\frac{1}{2\omega\chi^{2}}\nabla^{2}_{\theta}F=2\omega \Phi F.
\end{align}
where we used the polar coordinates $(\chi, \theta, \phi)$ with the source of GWs at the origin.
In this expression, $\nabla_{\theta}^{2}$ is a two-dimensional Laplace operator on two sphere defined as $\nabla_{\theta}^{2}=\partial^{2}/\partial \theta^{2}+\tan^{-1}{\theta}\partial/\partial\theta+\sin^{-2}\theta\partial^{2}/\partial\phi^{2}$.
Up to this point, we are using the coordinate system in which the source is at the origin; however, it is more common to switch the location of the observer and GW source as mentioned in \citep{Nakamura.Deguchi1999jan}.
In addition to switching the observer and source, \citep{Nakamura.Deguchi1999jan} uses the flat approximation in which the waves reaching the observer are assumed to be confined to the region where $\theta\ll 1$.
Under this approximation, it is appropriate to set $\sin{\theta}\sim \theta$ and regard $\bm \theta=\theta(\cos{\phi},\sin{\phi})$ as a two-dimensional vector on a flat plane. 
In the present paper, we follow the same coordinate system in \citep{Nakamura.Deguchi1999jan} by placing the GW source at $(\chi_{s}, \bm \theta_{s})$ as shown in Fig.~\ref{fig: GW lens}.

Note that the solution to Eq.~(\ref{eq: Amplification factor F}) is generally nonlinear in $\Phi$ even though $|\Phi|\ll1$ is assumed. 
This is because the gravitational potential induces the Shapiro time delay and its effect manifests itself as a phase in the exponent. Technically, the effect of the higher-order terms in $\Phi$ in Eq.~(\ref{eq: Amplification factor F}) appears as higher order terms in $\mathcal{O}(\Phi\omega\chi_{s})$, where $\chi_{s}$ is the distance from the source to the observer, and this is not necessarily small even if $\Phi\ll1$ (see Appendix \ref{app: derv SK}).
Physically, this implies that the phase change of GWs during propagation from the source to the observer becomes significant and leads to complex nonlinear interference effects.
For this reason, it is necessary to compute this equation to full order in $\Phi$ to obtain the comprehensive lensing effects.

On the other hand, in the context of weak lensing, it is assumed that $\Phi$ is sufficiently small that the expansion of $F$ in $\Phi$ up to first order provides a reasonable estimate of the true value of the amplification factor.
This approximation (i.e., the Born approximation) is primarily used to probe the small-scale power spectrum \citep{Takahashi2006jun,Oguri.Takahashi2020sep}.
In the Born approximation, the real and imaginary parts of the amplification factor are defined as the magnification $K$ and phase modulation $S$, which are functions of the GW frequency $\omega$, the line of sight distance $\chi_{s}$ to the source, and the angular coordinate $\bm \theta_{s}$ perpendicular to the line of sight.
In this definition, $K$ is related to the absolute value of $F$, and $S$ is interpreted as the argument of $F$.

Following the Born approximation, 
a systematic scheme to handle post-Born corrections was formulated by \cite{Mizuno.Suyama2023aug}, which introduced a new definition of $S$ and $K$ as $F(\omega)\equiv e^{K(\omega)}e^{iS(\omega)+i\omega \Delta t_{s}}$. 
Here, $\omega\Delta t_{s}$ is a shift of the phase due to the Shapiro time delay and is separated from $S(\omega)$ as the Shapiro time delay is not directly observable.
In the post-Born approximation, $K$ and $S$ are computed to second order in $\Phi$ as follows (Appendix \ref{app: derv SK} for derivation):
\begin{align}
    \label{eq: phase S 1st order}
    S^{(1)}=&-2\omega
    \int_{0}^{\chi_{s}}d\chi 
    \left[\cos{\left[\frac{W(\chi,\chi_{s})\nabla^{2}_{\theta}}{2\omega}\right]}-1\right]
    \Phi,
    \\
    \label{eq: phase S 2nd order}
    S^{(2)}=&-2\omega
    \int_{0}^{\chi_{s}}\frac{d\chi}{\chi^{2}}
    \int_{0}^{\chi}d\chi_{1}
    \int_{0}^{\chi}d\chi_{2} \notag
    \\
    &\times
    \left[\cos{\left[\frac{(W\nabla)^{(2)}}{2\omega}\right]}-1\right](\nabla_{\theta1}
    \Phi_{1}\cdot\nabla_{\theta2}\Phi_{2}),
    \\
     \label{eq: mag K 1st order}
    K^{(1)}=&2\omega
    \int_{0}^{\chi_{s}}d\chi 
    \sin{\left[\frac{W(\chi,\chi_{s})\nabla^{2}_{\theta}}{2\omega}\right]}
    \Phi,
    \\
    \label{eq: mag K 2nd order}
    K^{(2)}=&2\omega
    \int_{0}^{\chi_{s}}\frac{d\chi}{\chi^{2}}
    \int_{0}^{\chi}d\chi_{1}
    \int_{0}^{\chi}d\chi_{2} \notag
    \\
    &\times
    \sin{\left[\frac{(W\nabla)^{(2)}}{2\omega}\right]}
    (\nabla_{\theta1}\Phi_{1}\cdot\nabla_{\theta2}\Phi_{2}),
\end{align}
where $\Phi_{1(2)}=\Phi(\chi_{1(2)},\bm \theta),$ $W(\chi,\chi_{s})=1/\chi-1/\chi_{s}$ and,
\begin{align}
    \label{eq: 2nd order W nabla }
    (W\nabla)^{(2)}=&
    W(\chi,\chi_{s})\nabla^{2}_{\theta 12}
    +W(\chi_{1},\chi)\nabla^{2}_{\theta1}
    +W(\chi_{2},\chi)\nabla^{2}_{\theta2}.
\end{align}
In addition, the Shapiro time delay is given in the same manner up to second order as follows:
\begin{align}
    \label{eq: Shapi 1st order}
    \Delta t_{s}^{(1)}=&-2
    \int_{0}^{\chi_{s}}\Phi d\chi 
    ,
    \\
    \label{eq: Shapi 2nd order}
    \Delta t_{s}^{(2)}=&-2
    \int_{0}^{\chi_{s}}\frac{d\chi}{\chi^{2}}
    \int_{0}^{\chi}d\chi_{1}
    \int_{0}^{\chi}d\chi_{2} 
   \nabla_{\theta1}
    \Phi_{1}\cdot\nabla_{\theta2}\Phi_{2}.
\end{align}
Note that the derivatives are taken with respect to $\bm \theta$ with the operator $\nabla_{\theta 12}^{2}$ acting on both $\Phi_{1}$ and $\Phi_{2}$, while $\nabla_{\theta 1(2)}^{2}$ only acts on $\Phi_{1(2)}$. 
Also, the derivative operators involving the trigonometric functions (e.g., $\cos{\left[(W(\chi,\chi_{s})\nabla^{2}_{\theta})/(2\omega)\right]}$) are defined through the Fourier transform.  
In other words,  for arbitrary functions $F(\bm x)$ of a differential operator $\bm x$ and $f(\chi,\bm \theta)$, $F(\nabla_{\theta})f(\chi,\bm \theta)=\int \bm d\bm k_{\perp}/(2\pi)^{2} F(i\chi \bm k_{\perp})\tilde{f}(\chi,\bm k_{\perp})e^{i\bm k_{\perp} \cdot\chi \bm \theta }$ where $\tilde{f}(\chi,\bm k_{\perp})$ is the Fourier transform of $f(\chi,\bm\theta)$ with respect to $\chi \bm\theta$ defined as $\tilde{f}(\chi,\bm k_{\perp})=\int \chi^{2}d\bm \theta f(\chi,\bm \theta)e^{-i\chi \theta\cdot\bm k_{\perp}}$ and $\bm k_{\perp}$ is a wave vector on the two sphere.
In addition, the integral is taken along the straight line connecting the source and observer.
In these expressions, the first order terms $S^{(1)}$ and $K^{(1)}$ are the Born approximation, where $K^{(1)}$ reduces to the linear order convergence $\kappa$ in geometric optics in the high-frequency limit.

As is common in the context of weak lensing in geometric optics, the lensing signals are treated as random variables and the averages $\braket{\cdots}$ of these quantities are considered.
Using the power spectrum of the gravitational potential $\Phi$ combined with the Limber approximation, it is shown that, the following is satisfied for arbitrary functions $F(\bm y)$ and $G(\bm y)$ of the two-dimensional differential operator $\bm y$:
\begin{align}
     \label{eq: 2 point correlation}
    &\braket{
    F(\nabla_{\theta1})\Phi_{1}
    G(\nabla_{\theta2})\Phi_{2}}\notag
    \\
    &=
    \delta^{D}(\chi_{1}-\chi_{2})
    \int\frac{d^{2}\bm k_{\perp}}{(2\pi)^{2}}
    F(i\chi_{1}\bm k_{\perp})
    G(-i\chi_{1}\bm k_{\perp})
    P_{\Phi}(k_{\perp},\chi_1).
\end{align}
Here, we introduced the power spectrum of the gravitational potential $P_{\phi}(k,\chi)$ defined as
\begin{align}
    \braket{\tilde{\Phi}(\bm k_{1},\chi)\tilde{\Phi}(\bm k_{2},\chi)}
    =&(2\pi)^{3}\delta^{D}(\bm k_{1}+\bm k_{2})P_{\Phi}(k_{1},\chi),
\end{align}
where $\tilde{\Phi}(\bm k,\chi)$ is the Fourier transform of the gravitational potential and $\delta^{D}(\bm k)$ is the delta function.
With this relation and Eqs.~(\ref{eq: phase S 1st order})--(\ref{eq: mag K 2nd order}), we obtain
\begin{widetext}
\begin{align}
    \label{eq: S}
    \braket{S}=&2\omega
    \int_{0}^{\chi_{s}} 
    \frac{d\chi}{\chi^{2}}
    \int_{0}^{\chi}d\chi_{1} 
    \chi_{1}^{2}
    \int\frac{d^{2}\bm k_{\perp}}{(2\pi)^{2}} 
    k_{\perp}^{2}
    \left(1-\cos{\left[\frac{(\chi-\chi_{1})\chi_{1}}{\chi\omega}k_{\perp}^{2}\right]}\right)
    P_{\Phi}(k_{\perp},\chi_1), 
    \\
    \label{eq: K}
    \braket{K}=&-2\omega
    \int_{0}^{\chi_{s}} 
    \frac{d\chi}{\chi^{2}}
    \int_{0}^{\chi}d\chi_{1}\chi_{1}^{2}
    \int\frac{d^{2}\bm k_{\perp}}{(2\pi)^{2}} 
    k_{\perp}^{2}
    \sin{\left[\frac{(\chi-\chi_{1})\chi_{1}}{\chi\omega}k_{\perp}^{2}\right]}
    P_{\Phi}(k_{\perp},\chi_1),
\end{align}
for the averages, and, we obtain the following:
\begin{align}
    \label{eq: SS}
    \braket{S^{2}}=&4\omega^{2}
    \int_{0}^{\chi_{s}}d\chi
    \int\frac{d^{2}\bm k_{\perp}}{(2\pi)^{2}}
    \left[1-\cos{\left(\frac{(\chi_{s}-\chi)\chi}{2\chi_{s}\omega}k_{\perp}^{2}\right)}\right]^{2}
     P_{\Phi}(k_{\perp},\chi),
    \\
    \label{eq: KK}
    \braket{K^{2}}=&4\omega^{2}
    \int_{0}^{\chi_{s}}d\chi
    \int\frac{d^{2}\bm k_{\perp}}{(2\pi)^{2}}
    \sin^{2}{\left[\frac{(\chi_{s}-\chi)\chi}{2\chi_{s}\omega}k_{\perp}^{2}\right]}
     P_{\Phi}(k_{\perp},\chi),
    \\
    \label{eq: SK}
    \braket{SK}=&-4\omega^{2}
    \int_{0}^{\chi_{s}}d\chi
    \int\frac{d^{2}\bm k_{\perp}}{(2\pi)^{2}}
    \sin{\left[\frac{(\chi_{s}-\chi)\chi}{2\chi_{s}\omega}k_{\perp}^{2}\right]}
    \left(
    1-\cos{\left[\frac{(\chi_{s}-\chi)\chi}{2\chi_{s}\omega}k_{\perp}^{2}\right]}\right)
     P_{\Phi}(k_{\perp},\chi),
\end{align}
\end{widetext}
for the variances and the correlation between $S$ and $K$.

In these expressions, the scale at which the argument of the trigonometric functions becomes order unity provides a rough scale at which GWs are particularly sensitive.
This particular scale is referred to as the Fresnel scale $r_{F}=\sqrt{\chi(\chi_{s}-\chi)/\chi_{s}\omega}$.
In the context of lensing of GWs, the Fresnel scale is expressed as follows \citep*{Macquart2004aug,Takahashi2006jun}:
\begin{align}
    \label{eq:Fresnel scale}
    r_{F}\sim120\mathrm{pc}\left(\frac{f}{\rm mHz}\right)^{-1/2}\left[\frac{\chi(\chi_{s}-\chi)/\chi_{s}}{10\rm Gpc}\right]^{1/2},
\end{align}
where $f=\omega/2\pi$.
The Fresnel scale varies with the GW frequency $\omega$; thus, measuring the frequency dependence of $\braket{S^{2}}$,$\braket{K^{2}}$, $\braket{SK}$,$ \braket{S}$, and $\braket{K}$ is expected to be a unique probe for density fluctuations at scales as small as $k\simeq10^{6}-10^{8}\rm Mpc^{-1}$ for $f=10-1000$~Hz \citep{Takahashi2006jun,Oguri.Takahashi2020sep,Mizuno.Suyama2023aug}. 
Since the frequency dependence becomes relevant in the following discussion, the notations $S_{\omega}$ and $K_{\omega}$ are used to indicate the frequency dependence of each lensing signal (e.g., $\braket{S_{\omega}}=\braket{S(\omega)}$).

\section{Consistency relations}\label{sec: consistency}
The expressions for the averages (Eqs.~(\ref{eq: S}) and (\ref{eq: K})) can be simplified by exchanging the order of the integral as $\int_{0}^{\chi_{s}}d\chi \int_{0}^{\chi}d\chi_{1}$ $\to$ $\int_{0}^{\chi_{s}}d\chi_{1}\int_{\chi_{1}}^{\chi_{s}}d\chi$.
Then it is straightforward to obtain the following:
\begin{widetext}
\begin{align}
     \label{eq: S-2}
    \braket{S_{\omega}}=&2\omega^{2}
    \int_{0}^{\chi_{s}}d\chi_{1}
    \int\frac{d^{2}\bm k_{\perp}}{(2\pi)^{2}}
    \left\{
    \left(\frac{(\chi_{s}-\chi_{1})\chi_{1}}{\chi_{s}\omega}k_{\perp}^{2}\right)
    -
    \sin{\left(\frac{(\chi_{s}-\chi_{1})\chi_{1}}{\chi_{s}\omega}k_{\perp}^{2}\right)}
    \right\}
    P_{\Phi}(k_{\perp},\chi_{1}), 
    \\
    \label{eq: K-2}
    \braket{K_{\omega}}=&-4\omega^{2}
    \int_{0}^{\chi_{s}}d\chi_{1}
    \int\frac{d^{2}\bm k_{\perp}}{(2\pi)^{2}}
    \sin^{2}{\left[\frac{(\chi_{s}-\chi_{1})\chi_{1}}{2\chi_{s}\omega}k_{\perp}^{2}\right]}
     P_{\Phi}(k_{\perp},\chi_{1}).
\end{align}
\end{widetext}
By comparing these expressions with Eqs.(\ref{eq: SS})--(\ref{eq: SK}), we can readily find that the following consistency relations, which are accurate up to second order in $\Phi$:
\begin{align}
    \label{eq: consistencyK}
    \braket{K_{\omega}^{2}}+\braket{K_{\omega}}=&0,
    \\
    \label{eq: consistencyS}
    \braket{S_{\omega}}-\frac{1}{2}\braket{S_{2\omega}}
    =&-\braket{S_{\omega}K_{\omega}}.
\end{align}
Note that, to the best of our knowledge, these consistency relations have not been previously reported.
These relations involve the averages of $S$ and $K$, which vanish in the Born approximation and only appear at the level of the post-Born approximation.
The discovery of these relations was possible by considering the post-Born approximation within the wave optics framework.
In addition, the consistency relations derived here can provide new insight into an existing consistency relation derived by \cite{Inamori.Suyama2021sep}, which is explicitly expressed as follows: 
\begin{align}
    \label{eq: Inamori}
    \braket{S_{\omega}^{2}}+\braket{K_{\omega}^{2}}=\braket{K_{2\omega}^{2}}.
\end{align}
We observe that this consistency relation can be merged with Eq.~(\ref{eq: consistencyS}) as a single consistency relation for a complex-valued quantity using Eq.~(\ref{eq: consistencyK}).
By combining Eqs.~(\ref{eq: consistencyS}) and (\ref{eq: Inamori}), we obtain the following equivalent consistency relation:
\begin{align}
    \braket{K_{\omega}+iS_{\omega}}-\frac{1}{2}\braket{K_{2\omega}+iS_{2\omega}}
    =&-\frac{1}{2}\Braket{\left(K_{\omega}+iS_{\omega}\right)^{2}}.
\end{align}
In the following, we demonstrate that these relations can be derived as the weak lensing limit of more general relations that are accurate to full-order in $\Phi$.
In particular, we demonstrate that the consistency relation~(\ref{eq: consistencyK}) arises from the energy conservation of GWs.
A similar relation for the convergence $\kappa$ ($\braket{\kappa^{2}}=-2\braket{\kappa}$) \citep{Takahashi.Oguri.ea2011jan,Kaiser.Peacock2016feb} is derived under the photon number conservation in geometric optics \citep{Weinberg1976aug}; however, the discussion based on energy conservation is more general because it includes both geometric and wave optics.
On the other hand, the consistency relations~(\ref{eq: consistencyS}) and (\ref{eq: Inamori}) appear to be attributed to the Shapiro time delay, which is discussed in the subsection \ref{subsec:aveF}.

\subsection{Ensemble average}
The main results presented above, i.e., Eqs.~(\ref{eq: consistencyK}) and (\ref{eq: consistencyS}), are based on the computation of the average $\braket{\cdots}$ without paying particular attention to its meaning. 
However, it is important to revisit the meaning of the average to ensure a precise understanding of its implications, particularly in relation to the energy conservation law. 
In addition, it is also essential for determining how the average should be practically taken in future experimental settings.

The average considered to this point in this paper is referred to as the ensemble average \citep{Ellis.Maartens.ea2012,Breton.Fleury2021nov}, which hypothetically assumes the existence of multiple universes, each with different matter density configurations. 
In this scenario, we can compute the lensing signal $X(\chi_{s},\bm \theta_{s})$ (e.g., $S, K$ in wave optics and $\kappa, \gamma$ in geometric optics) by considering the GW(or light) signals from the same source at a fixed distance $\chi_{s}$ in each realization. 
Note that since $X$ describes the lensing effect, it does not depend on the physical property of the source.
The ensemble average is then obtained by taking the average value of $X$ over the ensemble of universes.
This is the original meaning of the ensemble average that we implicitly assumed in the previous discussion. 
In cosmology, it is presumed that the universe is statistically homogeneous and isotropic, meaning that the average of all realizations of universes is homogeneous and isotropic, even if each individual realization is not necessarily so.
This implies that the spatial derivative of $\braket{X}$ with respect to the true location of the source always vanishes; thus, we obtain the following:
\begin{align}
    \label{eq:nab<>=0}
    \nabla_{\theta}\braket{X(\chi_{s},\bm \theta_{s})}=0.
\end{align}
However, in reality, we only have access to a single realization of the universe, thereby making the true ensemble average unattainable.
Therefore, it becomes necessary to replace the ensemble average with a statistically computable averaging process.
In a statistically homogeneous and isotropic universe, one can find that the ensemble average is approximated by the average over the observers which represents the mean value of $X$ measured by a number of observers uniformly populated on the surface of a sphere with radius $\chi_{s}$ surrounding a single source.
This allows us to rewrite $\braket{X}$ as follows:
\begin{align}
    \label{eq:<>=AvOb}
    \braket{X}=
    \frac{1}{4\pi}
    \int X(\bm \theta)d\Omega
\end{align}
Note that $\bm \theta$ is the location of the observers on the surface of a sphere with radius $\chi_{s}$ surrounding the source.

However, we can only observe the source from the Earth; thus, it remains unfeasible to directly compute the average over the observers. 
In practice, $\braket{X}$ is taken as the average over the sources, which represents the mean value of $X$ computed from various sources located at the same fixed distance $\chi_{s}$.
It is obtained by simply summing all lensing signals $X$ from the sources at $\chi_{s}$ and dividing the sum by the number of the sources.
As long as each individual source is fully resolved, the average over the sources can be identified as the ensemble average.
In our context, we focus on a GW signal from binary systems where each individual source can be identified; thus, the ensemble averages of the lensing signals derived in the previous section ($\braket{S^{2}},\braket{K^{2}}$, etc.) should be taken as the average over the sources.

It is important to emphasize that $\braket{X}$, which, as discussed above, should not be confused with the average over the apparent directions of the sources within the framework of geometric optics.
The average over the directions is another approach commonly used in cosmology to compute the average of $X$ \citep{Kibble.Lieu2005oct,Fleury.Clarkson.ea2017mar} and is computed in a practical manner by dividing the celestial sphere into small patches with equal area and averaging $X$ over these patches.

The difference between the average over the sources and the average over the directions may seem subtle and indeed can be disregarded within the Born approximation (i.e., the first-order approximation of $X$).
However, when the higher-order terms are taken into account, making the distinction between these two becomes crucial, and failure to do so results in erroneous outcomes \citep{Bonvin.Clarkson.ea2015jun,Kaiser.Peacock2016feb}.

\subsection{Energy conservation}
Before delving into the main discussion, it is important to consider the meaning of the energy of GWs.
Although defining the energy of GWs is not as simple as the case of electromagnetic waves, it is still possible to assign energy to GWs as a conserved quantity when there is a clear separation of scales \citep{Isaacson1968feba}.
In the context of gravitational lensing, there are two types of metric perturbations: the gravitational potential $\Phi$ due to the presence of matter inhomogeneity and the metric perturbation caused by the GWs themselves. 
Here we assume that the wavelength of GWs is much shorter than the typical curvature radius of the gravitational potential; thus, the metric perturbation associated with GWs can be separated from the background metric.
As a result, we can treat GWs as a classical field just like any other fields living in an inhomogeneous universe described in Eq.~(\ref{eq:FRW metric}).
This approach enables us to identify a conserved quantity corresponding to the energy of GWs \citep{Maggiore2018apr}.

With this in mind, we can observe that Eq.~(\ref{eq: wave equation}) is essentially a wave equation with the lensing effect included as an interaction between GWs and the gravitational potential $\Phi$.
Thus, it can be rewritten as follows:
\begin{align}
    \label{eq:ConsvEngy}
    \frac{\partial }{\partial t}
    \left(\frac{1}{2}(\nabla \phi)^{2}+\frac{1}{2}\dot{\phi}^{2}-2\Phi\dot{\phi}^{2}\right)
    =
    -\nabla \cdot(-\dot{\phi}\nabla\phi).
\end{align}
Now, let us consider the volume integral over the region $V$ whose surface is denoted as $S$. 
Here, since the energy of GWs in a certain region is given by Eq.~(\ref{ap: energy}) in Appendix \ref{app: Energy}, 
we can connect Eq.~(\ref{eq:ConsvEngy}) with the energy conservation law by taking the time average of Eq.~(\ref{eq:ConsvEngy})\footnote{The time average is defined as $\braket{A}_{t}=(1/T)\int_{t}^{t+T}dt' A(t')$ for an arbitrary time-dependent quantity $A(t)$. Note that $\braket{A}_{t}$ is still a function of time.} in addition to the spatial integral.
Then, Eq.~(\ref{eq:ConsvEngy}) can be rewritten as follows:
\begin{align}
    \label{eq:ConsvEngyInteg}
    \frac{d E}{d t}
    =-\frac{1}{16\pi G}\int_{t}^{t+T}\frac{dt'}{T}\int_{S} dS \bm n\cdot (-\dot{\phi}\nabla\phi),
\end{align} 
where $\bm n$ is a unit normal vector at each point on $S$ and $T$ is a range of time average, which is taken sufficiently longer than the period of GWs.
From this expression, it is clear that the left-hand side represents the average rate at which the total energy in the region $V$ varies, and the right-hand side represents the average energy flow going into $V$. 
Thus, when the sign of the right-hand side is flipped, it is interpreted as the the energy going out from $V$.

Suppose the GW source is at the origin of the coordinate and $\phi$ is the superposition of the different frequency modes:
\begin{align}
    \phi(\bm x,t)
    =
    \int\frac{d\omega}{2\pi}
    \frac{e^{i\omega\chi-i\omega t}}{\chi}h(\omega)F(\omega, \bm x),
\end{align}
where $h(\omega)$ is the Fourier transform of the original waveform.
Next, we consider a sphere with the radius $\chi$. 
By taking the volume integral over this region and the time average, we obtain the following:
\begin{align}
    \label{eq:EnergyConsvCurrent}
    &\int_{t}^{T+t}\frac{dt'}{T}\int_{S} dS \bm n\cdot (-\dot{\phi}\nabla\phi)\notag
    \\
    =&\frac{1}{T}\int \frac{d\omega}{2\pi} ~\omega^{2}|h(\omega)|^{2}\int d\Omega F(\omega)F^{*}(\omega),
\end{align}
where $\Omega$ is a solid angle.
Therefore, we obtain the following:
\begin{align}
    \frac{d E}{d t}
    =&
    -\frac{1}{16\pi G T} \int \frac{d\omega}{2\pi}~\omega^{2}|h(\omega)|^{2}\int d\Omega F(\omega)F^{*}(\omega).
\end{align}
When the GW source is completely confined in the region $V$ and there are no objects in $V$ that absorb or produce GWs, then the right-hand side, especially $\int d\Omega F(\omega)F^{*}(\omega)$, becomes independent of the radius of a sphere $\chi$ surrounding the source.
In addition, the left-hand side is independent of the matter distribution in the region $V$ assuming that the gravitational potential does not significantly change over time; thus, the right-hand side is also not subject to this dependence. 
Given that $F=1$ when there are no lensing effects, $\int d\Omega FF^{*}$ needs to be normalized as follows:
\begin{align}
    1=\frac{1}{4\pi}\int d\Omega FF^{*}.
\end{align}
The right-hand side is the average of $FF^{*}$ over the observers, and it is identical to both the ensemble average and the average over the sources; thus, we obtain the following relation for the average of the absolute square of $F$:
\begin{align}
    \label{eq:FF=1}
    \Braket{FF^{*}}=1.
\end{align}
In our notation, the magnification $K$ and the phase modulation $S$ are defined as $F=e^{K+iS+i\omega \Delta t_{s}}$, which allows us to rewrite the energy conservation condition as $\Braket{e^{2K}}=1$.
In a weak lensing regime, $K$ is sufficiently smaller than unity and the Taylor expansion of $e^{2K}$ up to second order in $K$ provides $e^{2K}=1+2K+2K^{2}+\mathcal{O}(K^{3})$.
From this expression, it is clear that, up to second order in $\Phi$, $\Braket{K^{2}} +\Braket{K}$=0 needs to hold.
One noteworthy aspect of the relation $\braket{FF^{*}}=1$ is its generality.
It is the full-order result and does not assume any specific distribution of matter.

\subsection{Average of amplification factor}\label{subsec:aveF}
In the following, we explain a more general way to derive the consistency relations~(\ref{eq: consistencyS}) and (\ref{eq: Inamori}).
The physical interpretation of these relations may not be as clear as the consistency relation associated with energy conservation; however, 
they can still be derived from a more general, full-order condition, similar to how $\Braket{K}=-\braket{K^{2}}$ is directly derived from $\braket{FF^{*}}=1$.

By observing Eq.~(\ref{eq: Amplification factor F}), it is clear that the expression takes the same form as the Schr\"{o}dinger equation with time-varying mass.
Therefore it is possible to obtain the formal solution to this equation using the path integral method, as presented by \cite{Nakamura.Deguchi1999jan}:
\begin{align}
    \label{eq: Fpathintg}
    F(\omega, \chi_{s},\bm\theta_{s})
    =&
    \int \mathcal{D}[\bm \theta(\chi)]\notag
    \\
    &\times
    \exp{
    \left[
    i\int_{0}^{\chi_{s}}
    \left(
    \frac{1}{2}\omega\chi^{2}\left|\frac{d\bm\theta(\chi)}{d\chi}\right|^{2}
    -2\omega\Phi (\chi,\bm\theta(\chi))
    \right)
    d\chi
    \right]
    },
\end{align}
where the normalization factor is absorbed in $ \mathcal{D}[\bm \theta(\chi)]$ and is determined to satisfy $F=1$ when $\Phi=0$. 
Now, we consider taking the ensemble average of this expression.
When the ensemble average is taken, the only random variable that appears in this expression is $\Phi$.
Thus, $\braket{F}$ is given as follows:
\begin{align}
    \braket{F}
    =&
    \int \mathcal{D}[\bm \theta(\chi)]
    \exp{
    \left[
    i\frac{1}{2}\omega\int_{0}^{\chi_{s}}
    \chi^{2}\left|\frac{d\bm\theta(\chi)}{d\chi}\right|^{2}d\chi
    \right]
    }\notag
    \\
    &\times
    \Braket{e^{
    -2i\omega\int_{0}^{\chi_{s}} \Phi (\chi,\bm\theta(\chi)) d\chi
    }}.
\end{align}
Here, the computation of the $n$ point correlation function $\Braket{\Phi(\chi_{1},\bm \theta(\chi_{1}))\cdots\Phi(\chi_{n},\bm \theta(\chi_{n}))}$ is required to obtain $\Braket{\exp{\left[-2i\omega\int_{0}^{\chi_{s}} \Phi (\chi,\bm\theta(\chi)) d\chi\right]}}$. 
By considering the spatial homogeneity and the assumption that the potential $\Phi$ evaluated at different $\chi$ is uncorrelated (the Limber approximation), we obtain $\Braket{\exp{\left[-2i\omega\int_{0}^{\chi_{s}} \Phi (\chi,\bm\theta(\chi)) d\chi\right]}}=\Braket{\exp{\left[-2i\omega\int_{0}^{\chi_{s}} \Phi (\chi,\bm\theta_{s}) d\chi\right]}}$\footnote{Because 
$\Braket{\Phi(\chi_{1},\bm \theta(\chi_{1}))\cdots\Phi(\chi_{n},\bm \theta(\chi_{n}))}
=\delta^{D}(\chi_{1}-\chi_{2})\cdots\delta^{D}(\chi_{n-1}-\chi_{n})\Braket{\Phi(\chi_{1},\bm \theta(\chi_{1}))\cdots\Phi(\chi_{1},\bm \theta(\chi_{1}))}_{\perp}
=\Braket{\Phi(\chi_{1},\bm \theta_{s})\cdots\Phi(\chi_{1},\bm \theta_{s})}$, where $\braket{\cdots}_{\perp}$ indicates the ensemble average on the plane perpendicular to the line of sight.}.
Then, $\braket{F}$ is further simplified as follows:
\begin{align}
    \label{eq: Fave}
    \braket{F( \chi_{s},\bm\theta_{s})}
    =
    \Braket{
    \exp{\left(
    -2i\omega\int_{0}^{\chi_{s}} \Phi (\chi,\bm\theta_{s}) d\chi
    \right)
    }}
    =&\Braket{e^{i\omega \Delta t_{s}^{(1)}}}.
\end{align}
This is a surprisingly simple relation that is accurate to full-order.
Here, $F$ is written as $F=e^{K(\omega)}e^{iS(\omega)+i\omega \Delta t_{s}}$; thus, this expression can be formally expanded in $\Phi$ as follows:
\begin{align}
    &1+\Braket{K+S+\omega \Delta t_{s}}
    +\frac{1}{2} 
    \Braket{(K+iS+i\omega \Delta t_{s})^{2}}+\mathcal{O}(\Phi^{3})
    \notag
    \\
    =&
    1-\frac{\omega^{2}}{2}\Braket{(\Delta t_{s}^{(1)})^{2}}+\mathcal{O}(\Phi^{3}).
\end{align}
From this relation and Eqs.~(\ref{eq: phase S 1st order})--(\ref{eq: Shapi 2nd order}), we obtain the following expression up to second order in $\Phi$:
\begin{align}
    \label{eq:SKcon}
    \braket{K_{\omega}+iS_{\omega}}-\frac{1}{2}\braket{K_{2\omega}+iS_{2\omega}}
    =&-\frac{1}{2}\Braket{\left(K_{\omega}+iS_{\omega}\right)^{2}}.
\end{align}
This is nothing more than Eqs.~(\ref{eq: consistencyS}) and Eq.~(\ref{eq: Inamori}).
In addition, the expressions of the consistency relations~(\ref{eq: consistencyS}) and (\ref{eq: Inamori}) are based partly on the Limber approximation, which was not assumed in the derivation of the consistency relation associated with energy conservation.

A notable difference between the consistency relations~(\ref{eq: consistencyS}) and (\ref{eq: Inamori}) and the one related to energy conservation~(\ref{eq: consistencyK}) is that  Eqs.~(\ref{eq: consistencyS}) and (\ref{eq: Inamori}) establish a nontrivial connection between the real and imaginary parts of the amplification factor (i.e., magnification $K$ and the phase modulation $S$ in weak lensing).
Here, we propose that this non-trivial relation arises from the Shapiro time delay.
As observed in Eq.~(\ref{eq: Fpathintg}), the amplification factor $F$ is obtained by the superposition of all waves traveling along various possible paths.
Since the presence of the gravitational potential in a particular region only induces a phase shift to the GWs passing through that area, the resulting $F$ undergoes changes in both the magnification and the phase modulation.
However, these changes are only due to constructive and destructive interference.
Thus, it is expected that there is a nontrivial connection between the magnification and the phase modulation, and it appears that this connection becomes apparent in the form of the consistency relations when the average is taken
\footnote{The nontrivial relation between the real and the imaginary parts of the amplification factor has been reported by \cite{Tanaka.Suyama2023aug} where the relation arises from the causality of GWs.}.

To obtain a more intuitive understanding of this nontrivial connection between the real and imaginary parts of $F$, we provide a simple toy model that demonstrates this effect.
Suppose two GWs with the same amplitude travel along different paths of equal length and arrive at the location of an observer.
Without any lensing objects, the amplification factor is $F=1$.
However, if one of the GWs passes through a region with nonzero gravitational potential $\Phi$ that extends over a length $\Delta\chi$, the resulting amplification factor can be written as follows:
\begin{align}
    F(\omega)=e^{K_{\omega}+iS_{\omega}}=\frac{1}{2}\left(1+e^{-2i\omega \Delta\chi \Phi}\right).
\end{align}
From this amplification factor, we obtain the expressions for the magnification $K_{\omega}$ and phase modulation $S_{\omega}$:
\begin{align}
    K_{\omega}=&\frac{1}{2}\ln{\left(\frac{1+\cos{(2\omega\Delta\chi \Phi)}}{2}\right)},
    \\
    S_{\omega}=&-\tan^{-1}{\left(\frac{\sin{(2\omega\Delta\chi \Phi)}}{1+\cos{(2\omega\Delta\chi \Phi)}}\right)}.
\end{align}
By expanding these expressions up to second order in $\Phi$, we can verify the following:
\begin{align}
   K_{\omega}+iS_{\omega}-\frac{1}{2}(K_{2\omega}+iS_{2\omega})
   =&-\frac{1}{2}(K_{\omega}+iS_{\omega})^{2}+\mathcal{O}(\Phi^{3}).
\end{align}
This relation is identical to Eq.~(\ref{eq:SKcon}) with the only difference being the absence of the averaging process. 
Therefore, it is reasonable to conclude that the Shapiro time delay is responsible for the origin of the consistency relation Eq.~(\ref{eq:SKcon}).

\subsection{Violation of consistency relations due to massive gravitons}\label{sec:massive graviton}
The consistency relations we derived above are satisfied irrespective of the shape of the matter power spectrum; thus, it is constructive to investigate the circumstances under which the consistency relations might be compromised, especially due to the violation of the fundamental physics principle we assumed rather than as a result of observational errors and biases.
Given that Eq.~(\ref{eq: consistencyK}) arises as a result of the energy conservation and Eqs.~(\ref{eq: consistencyS}) and (\ref{eq: Inamori}) (equivalently Eq.~(\ref{eq:SKcon})) are attributed to the Shapiro time delay under the assumption of GWs propagating at the speed of light, it is expected that Eq.~(\ref{eq: consistencyK}) remains to be satisfied whereas Eq.~(\ref{eq:SKcon}) may be violated if the speed of GW propagation is changed.
In order to see if this expectation is indeed correct, let us consider the case of massive gravitons since this is the simplest modification to GR to account for the change in the propagation speed of GWs. 
When the mass of a graviton $m$ is considered, the wave equation for the amplification factor $F$ is rewritten as follows:
\begin{align}
    i \frac{\partial F}{\partial \chi}+\frac{1}{2\omega\chi^{2}}\nabla^{2}_{\theta}F=2\omega \Phi F+\frac{m^{2}}{2\omega}F.
\end{align}
This expression indicates that a newly defined function $F'=F e^{\frac{im^{2}\chi_{s}}{2\omega}}=e^{K+i(S+\frac{m^{2}\chi_{s}}{2\omega})+i\omega\Delta t_{s}}$ satisfies the equation for a massless graviton (\ref{eq: Amplification factor F}).
As we have shown above, the magnification and phase modulation for a massless graviton satisfy Eq.~(\ref{eq:SKcon}), and in this case, the corresponding magnification and phase modulation are $K_{\omega}$ and $S_{\omega}+\frac{m^{2}\chi_{s}}{2\omega}$; thus, the modified version of the consistency relation when the mass of a graviton is included is obtained by simply replacing $S_{\omega}$ with $S_{\omega}+\frac{m^{2}\chi_{s}}{2\omega}$ as follows:
\begin{align}
    &\Braket{K_{\omega}+i\left(S_{\omega}+\frac{m^{2}\chi_{s}}{2\omega}\right)}-\frac{1}{2}\Braket{K_{2\omega}+i\left(S_{2\omega}+\frac{m^{2}\chi_{s}}{4\omega}\right)}\notag
    \\
    =&-\frac{1}{2}\Braket{\left(K_{\omega}+i\left(S_{\omega}+\frac{m^{2}\chi_{s}}{2\omega}\right)\right)^{2}}.
\end{align}
This modified version of the consistency relation implies that the deviation from Eq.~(\ref{eq: consistencyS}) is of the order $\frac{m^{2}\chi_{s}}{\omega}$ while the deviation from Eq.~(\ref{eq: Inamori}) is of the order $(\frac{m^{2}\chi_{s}}{\omega})^{2}$ when the mass of a graviton is considered.
Note that the consistency relation originating from the energy conservation (i.e., $\braket{K}=-\braket{K^{2}}$ or $\braket{FF^{*}}=\braket{e^{2K}}=1$) is unchanged even when the mass of a graviton is considered since $FF^{*}=\braket{e^{2K}}=1$ is unaffected by replacing $F$ with $F e^{\frac{im^{2}\chi_{s}}{2\omega}}$. 
Physically,  this is a consequence of the fact that energy conservation is still satisfied despite the presence of massive gravitons; thus, the associated consistency relation (\ref{eq: consistencyK}) also remains unchanged.

\subsection{Application}
The weak lensing signals $S$ and $K$ can be used to probe the small scale matter power spectrum \citep{Takahashi2006jun,Oguri.Takahashi2020sep}.
In order to achieve this, it is of critical importance to accurately extract correct $S$ and $K$ from the observational data. 
As suggested in \citep{Inamori.Suyama2021sep,Tanaka.Suyama2023aug}, consistency relations have the potential to serve as a means to verify the reliability of the lensing signal obtained from observational data. 
By confirming the satisfaction of the consistency relations, we can independently confirm the correctness of the observed lensing signals without assuming the shape of matter power spectrum, enabling us to use the lensing signals as probes for small-scale matter density fluctuations.
In addition, satisfaction of the consistency relations will confirm the validity of the general relativistic formulation of the lensing signals.
Conversely, any deviation from the consistency relations serves as a warning sign that the estimation of $S$ and $K$ may not have been performed correctly, which prevents incorrect results from being inferred from unreliable data.
While the primary objective of this paper is to present the new consistency relations and discuss their physical implications, it is worth providing a rough estimate of how well the presented consistency relations are satisfied under more realistic scenarios.

Therefore, we consider the feasibility of confirming the consistency relations 
following a similar method presented in \citep{Inamori.Suyama2021sep,Mizuno.Suyama2023aug}.
In practical situations, the average $\braket{\cdots}$ is taken as the average over the sources, which requires a number of GWs from various sources, e.g., binary black holes located at a fixed redshift.
However, in principle, it is impossible to collect a sufficient number of lensing signals from the sources with exactly the same redshift $z_{s}$; thus, it is necessary to redefine the average by allowing the inclusion of signals whose redshift falls within a range $z_{s}-\Delta z<z<z_{s}+\Delta z$.
The redshift dependence of the lensing signal $X(=S,K)$ suggests that the observed variance at $z_{s}+\Delta z$ is roughly given by $\braket{X(z_{s}+\Delta z)^{2}}=\braket{X(z_{s})^{2}}(1+\mathcal{O}(\Delta z))$ \citep{Takahashi2006jun,Oguri.Takahashi2020sep, Mizuno.Suyama2023aug}.
With this in mind, we define the estimators $\mathcal{E}_{A}$ and $\mathcal{E}_{B}$ as 
\begin{align}
    \mathcal{E}_{A}(\omega)=&
    \frac{1}{N}\sum_{i}
    (K_{i}^{2}(\omega, z_{i})+K_{i}(\omega, z_{i})),
    \\
    \mathcal{E}_{B}(\omega)
    =&
    \frac{1}{N}\sum_{i}
    \left(
    S_{i}(\omega,z_{i})-\frac{1}{2}S(2\omega,z_{i})
    +S_{i}(\omega,z_{i})K_{i}(\omega,z_{i})
    \right),
\end{align}
where $K_{i}$ and $S_{i}$ are assumed to contain independent Gaussian noise $n_{i}$ with zero mean and variance $1/\mathrm{SNR}^{2}$, where $\mathrm{SNR}$ is the signal-to-noise ratio of the detectors for a particular frequency of GWs. 
In addition, the products of the signals, e.g., $K_{i}^{2}(\omega, z_{i})$ and $S_{i}(\omega,z_{i})K_{i}(\omega,z_{i})$, are assumed to be computed using the two values obtained from different detectors with independent noise.
Under this assumption, we can immediately obtain $ \braket{\mathcal{E}_{A}}= \braket{\mathcal{E}_{B}}=0$.
Furthermore, under the assumptions of weak lensing, small $\Delta z$ ($\braket{X(z_{s}+\Delta z)^{2}}\sim\braket{X(z_{s})^{2}}$), and $|K|,|S|<1/\mathrm{SNR}$, we obtain, $ \braket{\mathcal{E}_{A}^{2}}^{1/2}\sim \braket{\mathcal{E}_{B}^{2}}^{1/2}\sim\frac{1}{\mathrm{SNR}}\frac{1}{\sqrt{N}}$, which provides the estimated fluctuations in $\mathcal{E}_{A}$ and $\mathcal{E}_{B}$.

The number of GW events expected to be observed per year within a redshift range $2.9<z_{s}<3$ can be estimated as $N\sim 10^{3}$ under the assumption that the merger rate at $z_{s}=3$ is $R=20~\mathrm{Gpc^{-3}yr^{-1}}$ \citep{Abbott.Abbott.ea2023mar}. 
In the $\rm SNR=50$ case, $\frac{1}{\mathrm{SNR}\sqrt{N}}\sim 6\times 10^{-4}$. 
Since $\sqrt{\braket{K^{2}}}\sim\mathcal{O}(10^{-2})$ and $\sqrt{\braket{S^{2}}}\sim\mathcal{O}(10^{-3})$ at $z_{s}\sim3$ and $f\sim 1$~Hz, in this scenario, the consistency relation (\ref{eq: consistencyK}) can be confirmed with an accuracy of approximately $\mathcal{O}(1)$\% of $\braket{K},\braket{K^{2}}$, and the consistency relation (\ref{eq: consistencyS}) can be confirmed with an accuracy of up to $\mathcal{O}(10)$\% of $\braket{S},\braket{SK}$.
Note that the value of the merger rate $R$ used here is an estimated value at a fiducial redshift $z=0.2$ (rather than $z=3$). 
Since $R$ is expected to take a larger value at higher redshift, the number of GW events we estimated might be moderately underestimated. 
Thus, in reality, the consistency relation can be even more tightly confirmed.

\section{Conclusion}\label{sec:cocl}
In this paper, we investigated the lensing of GWs with a particular focus on consistency relations.
In addition to the previously reported consistency relation \citep{Inamori.Suyama2021sep}, we have identified two additional consistency relations~(\ref{eq: consistencyK}) and (\ref{eq: consistencyS}) that are accurate in the weak lensing regime by directly computing the magnification $K$ and phase modulation $S$.
We have demonstrated that Eq.~(\ref{eq: consistencyK}) arises from the conservation of energy in GWs by demonstrating that Eq.~(\ref{eq: consistencyK}) is derived as the weak lensing limit of $\braket{FF^{*}}=1$.
In fact, $\braket{FF^{*}}=1$ holds to full order in $\Phi$ regardless of the shape or the correlation of the matter clumps. 
In addition, we have shown that the other consistency relations~(\ref{eq: consistencyS}) and (\ref{eq: Inamori}) can be also derived as the weak lensing limit of the average of the amplification factor $\braket{F}=\braket{e^{-2i\omega\int_{0}^{\chi_{s}}\Phi d\chi}}$, which is also accurate to full order in $\Phi$.
The analysis presented in this paper indicates that the consistency relations~(\ref{eq: consistencyS}) and (\ref{eq: Inamori}) appear to arise from the Shapiro time delay, which locally alters the phase of GWs.
This leads to interference effects and poses the nontrivial connection between $K$ and $S$, which becomes evident when the average is taken.
Finally, we have demonstrated that these consistency relations can be confirmed observationally given that sufficient SNR$\sim50$ is achieved.
Thus, we expect that they will provide independent verification of the correct observed lensing signals and enable us to properly probe matter density fluctuations at very small scales.

\section*{Acknowledgements}
This work was supported by the JSPS KAKENHI Grant Numbers JP19K03864 (TS), 
JP23K03411 (TS), JP22H00130 (RT) and JP20H05855 (RT).

\appendix
\section{Derivation of $S^{(1)}, S^{(2)}, K^{(1)},K^{(2)}$}\label{app: derv SK}
In this appendix, we review the derivation of $S^{(1)}, S^{(2)}, K^{(1)}$, and $K^{(2)}$ following the method developed in \citep{Mizuno.Suyama2023aug}. 
In \citep{Mizuno.Suyama2023aug}, a new variable $J$ defined as $F=e^{i\omega J}$ is used to include the effect beyond the Born approximation.
In order to clarify the reason for the necessity to introduce $J$, let us start by considering the expansion of $F$ in $\Phi$ and see why it is not the best way to investigate beyond the Born approximation. 
First, we rewrite Eq.~(\ref{eq: Amplification factor F}) as follows:
\begin{align}
     \label{eq: Amplification factor F app}
    \left(\frac{\partial }{\partial \chi}-\frac{i}{2\omega\chi^{2}}\nabla^{2}_{\theta}\right)F=-2i\omega \Phi F.
\end{align}
Note that $\bm \theta$ is a two-dimensional Cartesian coordinate vector, i.e., we are adopting the flat sky approximation ($\theta\ll1$) \citep{Nakamura.Deguchi1999jan}.
Under this assumption, $\nabla_{\theta}^{2}$ is a Laplace operator on a two-dimensional flat space defined as $\nabla_{\theta}^{2}=\partial^{2}/\partial \theta^{2}+\theta^{-1}\partial/\partial\theta+\theta^{-2}\partial^{2}/\partial\phi^{2}$. 
This equation can be formally solved by finding Green's function of the linear operator on the left-hand side of this expression. 
By definition, Green's function satisfies the following equation:
\begin{align}
    \label{eq: time dependent schrodinger equation}
    \left(\frac{\partial }{\partial \chi}-\frac{i}{2\omega \chi^{2}}\nabla_{\theta}^{2}\right)G(\chi-\chi',\bm \theta- \bm\theta')=\delta^{D}(\chi-\chi')\delta^{D}(\bm \theta-\bm \theta'),
\end{align}
which can be solved to be
\begin{align}
    &G(\chi-\chi',\bm \theta-\bm \theta')\notag
    \\
    &=\frac{i\omega}{2\pi}\frac{\chi \chi'}{(\chi-\chi')}\exp{\left[i\omega\frac{\chi \chi'}{2(\chi-\chi')}|\bm \theta-\bm \theta'|^{2}\right]}\Theta(\chi-\chi'),
\end{align}
where $\Theta(\chi)$ is a step function. 
Now, we assume that the observer is at $(\chi_{s},\bm\theta_{s})$ and the GW source is at the origin. 
Then, using this Green's function, $F(\chi_{s},\bm \theta_{s})$ is written as follows:
\begin{align}
    \label{eq: integ eq F}
    F(\chi_{s},\bm \theta_{s})=&\int d\chi \int d\bm \theta G(\chi_{s}-\chi,\bm \theta_{s}-\bm \theta)\left\{-2i\omega\Phi(\chi,\bm \theta) F\right\}
    \notag
    \\
    =&-2i\omega \int_{0}^{\chi_{s}} d\chi \int d\bm \theta \frac{i\omega}{2\pi}\frac{\chi_{s} \chi}{(\chi_{s}-\chi)}\notag
    \\
    &\quad \times\exp{\left[i\omega\frac{\chi_{s} \chi}{2(\chi_{s}-\chi)}|\bm \theta_{s}-\bm \theta|^{2}\right]}\Phi(\chi,\bm \theta) F(\chi,\bm \theta)
    \notag
    \\
    =&-2i\omega\int_{0}^{\chi_{s}} d\chi \exp{\left[i\frac{W(\chi,\chi_{s})\nabla_{\theta}^{2}}{2\omega}\right]}\Phi(\chi,\bm \theta_{s}) F(\chi,\bm \theta_{s})
\end{align}
where $W(\chi',\chi)=1/\chi'-1/\chi$.
From the second line to the third, the following formula was used:
\begin{align}
    \int d\bm y \exp{\left(i\omega|\bm y-\bm x_{0}|^{2}\right)}f(\bm y)=&\frac{i\omega}{\pi}\exp{\left(\frac{i\nabla^{2}_{\bm x_{0}}}{4\omega}\right)}f(\bm x_{0})
\end{align}
This formula can be verified by Fourier transforming $f(\bm y)$ and performing the Gaussian integral. 
Then, consider the expansion of $F$ in $\Phi$, namely, $F=1+F^{(1)}+F^{(2)}+\cdots$, where the zeroth order term is determined to be 1 since $F\to1$ when there is no lensing, i.e., $\Phi=0$.
By plugging $F=1+F^{(1)}+F^{(2)}+\cdots$ in Eq.~(\ref{eq: Amplification factor F app}), we obtain the following expressions up to second order in $\Phi$:
\begin{align}
    F^{(1)}=&-2\omega\int_{0}^{\chi_{s}} d\chi 
    \exp{\left[i\frac{W(\chi,\chi_{s})\nabla^{2}_{\theta}}{2\omega}\right]}\Phi
    (\chi,\bm\theta_{s}),
    \\
    F^{(2)}=&4\omega^{2}
    \int_{0}^{\chi_{s}} d\chi 
    \exp{\left[i\frac{W(\chi,\chi_{s})\nabla^{2}_{\theta}}{2\omega}\right]}\notag
    \\
    &\times
    \left[\Phi
    (\chi,\bm\theta_{s})\int_{0}^{\chi} d\chi' 
    \exp{\left[i\frac{W(\chi',\chi)\nabla^{2}_{\theta}}{2\omega}\right]}\Phi
    (\chi',\bm\theta_{s})\right].
\end{align}
From this expression, it is verified that $F^{(1)}\sim \mathcal{O}(\Phi\omega\chi_{s})$ and $F^{(2)}\sim \mathcal{O}((\Phi\omega\chi_{s})^{2})$. 
In fact, it can be shown through iteration that $F^{(n)}\sim\mathcal{O}((\Phi\omega\chi_{s})^{n})$.
Thus, the higher-order terms in $\Phi$ always appear as higher-order terms in $\Phi \omega \chi_{s}$ in the expansion of $F$.
Due to this property, there are two problems associated with expanding $F$, even though $F^{(n)}$ can be formally computed. 
Firstly, it is not clear how to obtain the geometric optics limit from this expression. Conceptually, geometric optics should be derived by taking $\omega\to \infty$, however, since $F^{(n)}\sim \mathcal{O}(\Phi\omega\chi_{s})$, it is required to compute $F$ to full order in $\Phi \omega\chi_{s}$ in order to accurately estimate $F$. 
The second problem is that the higher-order terms need to be converted into physical quantities such as the phase modulation and magnification through an additional process. 
For example, the phase modulation is obtained by computing the imaginary part of $\log{F}$; therefore, even if $F^{(n)}$ are obtained, we need to perform non-trivial calculations to obtain the correction terms to the phase modulation.

Having said that, it is possible to systematically compute the correction terms to the physical quantities by introducing a new variable $J$ defined as $F=e^{i\omega J}$  as suggested in \citep{Mizuno.Suyama2023aug}.
Using $J$, Eq.~(\ref{eq: Amplification factor F}) becomes
\begin{align}
    \label{eq: eq for J}
    \left(\frac{\partial }{\partial \chi}-\frac{i}{2\omega 
    \chi^{2}}\nabla_{\theta}^{2}\right)J=f(\chi,\bm \theta),
\end{align}
where $f(\chi,\bm \theta)=-2\Phi-(\nabla_{\theta}J)^{2}/(2\chi^{2})$. 
Following the same step, we can write this equation in the following form:
\begin{align}
    \label{eq: integ eq J}
    J(\chi_{s},\bm \theta_{s})=&\int d\chi \int d\bm \theta G(\chi_{s}-\chi,\bm \theta_{s}-\bm \theta)f(\chi,\bm \theta)\notag
    \\
    =&
    \int_{0}^{\chi_{s}} d\chi \exp{\left[i\frac{W(\chi,\chi_{s})\nabla_{\theta}^{2}}{2\omega}\right]}
    \left(
    -2\Phi-\frac{1}{2\chi^{2}}(\nabla_{\theta}J)^{2}
    \right)
\end{align}
Now, let us define $J^{(n)}$ as the components of $J$ proportional to the $n$th order of the gravitational potential, i.e., $J=J^{(1)}+J^{(2)}+\mathcal{O}(\Phi^{3})$.
By inserting $J=J^{(1)}+J^{(2)}+\mathcal{O}(\Phi^{3})$. into Eq.~(\ref{eq: integ eq J}) and equating the same order terms, $J^{(n)}$ can be formally obtained order by order.
Then, the expressions for $J^{(1)}$ and $J^{(2)}$ are given as follows:
\begin{align}
     \label{eq:J 1st order a }
    J^{(1)}(\chi_{s},\bm \theta_{s})=&\int_{0}^{\chi_{s}} d\chi 
    \exp{\left[i\frac{W(\chi,\chi_{s})\nabla^{2}_{\theta}}{2\omega}\right]}(-2\Phi
    (\chi,\bm\theta_{s})),
    \\
    \label{eq:J 2nd order a}
     J^{(2)}(\chi_{s},\bm\theta_{s})=&-\int_{0}^{\chi_{s}} d\chi 
     \exp{\left[i\frac{W(\chi,\chi_{s})\nabla^{2}_{\theta}}{2\omega}\right]}\frac{
     (\nabla_{\theta}J^{(1)}(\chi,\bm\theta_{s}))^{2}}{2\chi^{2}}.
\end{align}
In order to further simplify the expression for $J^{(2)}$, it is convenient to introduce several notions $\Phi_{1(2)}\equiv \Phi(\chi_{1(2)},\bm \theta_{s})$ and $\nabla_{\theta 1(2)}$ and $\nabla_{\theta 12}$. 
Here, $\nabla_{\theta12}$ are defined to act on both $\Phi_{1}$ and $\Phi_{2}$ whereas $\nabla_{\theta1(2)}$ only acts on $\Phi_{1(2)}$. 
Then, Eq.~(\ref{eq:J 2nd order a}) is rewritten as 
\begin{widetext}
\begin{align}
    J^{(2)}(\chi_{s},\bm\theta_{s})=&-2\int_{0}^{\chi_{s}} 
     \frac{d\chi}{\chi^{2}}
      \exp{\left[i\frac{W(\chi,\chi_{s})\nabla^{2}_{\theta12}}{2\omega}\right]}
      \int_{0}^{\chi}d\chi_{1}\int_{0}^{\chi}d\chi_{2} \nabla_{\theta1}\left( \exp{\left[i\frac{W(\chi_{1},\chi)\nabla^{2}_{\theta1}}{2\omega}\right]}\Phi_{1}\right)\cdot\nabla_{\theta2}\left( \exp{\left[i\frac{W(\chi_{2},\chi)\nabla^{2}_{\theta2}}{2\omega}\right]}\Phi_{2}\right)
\end{align}
\end{widetext}
In this notation, $\nabla_{\theta12}$ and $\nabla_{\theta1(2)}$ all commute with each other; thus, $J^{(2)}$ is given by the following expression:
\begin{align}
    \label{eq: J 2nd aa}
     J^{(2)}(\chi_{s},\bm\theta_{s})=&-2\int_{0}^{\chi_{s}} 
     \frac{d\chi}{\chi^{2}}\int_{0}^{\chi}d\chi_{1}\int_{0}^{\chi}d\chi_{2} 
     \exp{\left[i\frac{(W\nabla)^{(2)}}{2\omega}\right]}\notag
     \\
     &\times\nabla_{\theta1}\Phi_{1}\cdot\nabla_{\theta2}\Phi_{2}
\end{align}
where $
    (W\nabla)^{(2)}\equiv
    W(\chi,\chi_{s})\nabla^{2}_{\theta 12}
    +W(\chi_{1},\chi)\nabla^{2}_{\theta1}
    +W(\chi_{2},\chi)\nabla^{2}_{\theta2}.$
The  phase modulation $S$ and magnification $K$ are obtained from the relation $i\omega J = K+iS +i \omega \Delta t_{s}$, where the Shapiro time delay $\Delta t_{s}$ is given by $\Delta t_{s}=\lim_{\omega\to\infty}J(\omega)$ \citep{Mizuno.Suyama2023aug}.
Using Eq.~(\ref{eq:J 1st order a }), (\ref{eq: J 2nd aa}), and the definition of $K$ and $S$, the following expressions for $S^{(1)}, S^{(2)}, K^{(1)}$, and $K^{(2)}$ are obtained:
\begin{align}
    S^{(1)}=&-2\omega
    \int_{0}^{\chi_{s}}d\chi 
    \left[\cos{\left[\frac{W(\chi,\chi_{s})\nabla^{2}_{\theta}}{2\omega}\right]}-1\right]
    \Phi,
    \\
    S^{(2)}=&-2\omega
    \int_{0}^{\chi_{s}}\frac{d\chi}{\chi^{2}}
    \int_{0}^{\chi}d\chi_{1}
    \int_{0}^{\chi}d\chi_{2} \notag
    \\
    &\times
    \left[\cos{\left[\frac{(W\nabla)^{(2)}}{2\omega}\right]}-1\right](\nabla_{\theta1}
    \Phi_{1}\cdot\nabla_{\theta2}\Phi_{2}),
    \\
    K^{(1)}=&2\omega
    \int_{0}^{\chi_{s}}d\chi 
    \sin{\left[\frac{W(\chi,\chi_{s})\nabla^{2}_{\theta}}{2\omega}\right]}
    \Phi,
    \\
    K^{(2)}=&2\omega
    \int_{0}^{\chi_{s}}\frac{d\chi}{\chi^{2}}
    \int_{0}^{\chi}d\chi_{1}
    \int_{0}^{\chi}d\chi_{2} \notag
    \\
    &\times
    \sin{\left[\frac{(W\nabla)^{(2)}}{2\omega}\right]}
    (\nabla_{\theta1}\Phi_{1}\cdot\nabla_{\theta2}\Phi_{2}).
\end{align}
Note that in this expression, the geometric optics limit can be easily obtained by taking $\omega\to \infty$. Indeed, the magnification reduces to the weak lensing convergence in this limit (Appendix A of \citep{Mizuno.Suyama2023aug}).
In addition, higher-order corrections can be computed in the same way without going through any conversion processes.
Throughout this derivation, we consider that the GW source is located at the origin and the observer is at $(\chi_{s},\bm \theta_{s})$; however, as shown in \citep{Nakamura.Deguchi1999jan}, these expressions remain unchanged if we swap the location of the observer and GW source.

\section{Energy density of gravitational waves}\label{app: Energy}
Here, we provide a brief derivation of the energy density of GWs propagating in curved spacetime characterized by Eq.~(\ref{eq:FRW metric}).
When there is a clear separation between the metric components due to the background  $\overline{g}_{\mu\nu}$ (typical variation scale $L$) and highly oscillatory perturbations  $h_{\mu\nu}$ (typical wavelength $\lambda$), the total metric $g_{\mu\nu}$ is separated into two parts \citep{Isaacson1968feb}:
\begin{align}
    g_{\mu\nu}=\overline{g}_{\mu\nu}+h_{\mu\nu},
\end{align}
where $\overline{g}_{\mu\nu}$ is given by Eq.~(\ref{eq:FRW metric}).
The Einstein equations $R_{\mu\nu}-\frac{1}{2}g_{\mu\nu}R=8\pi G T_{\mu\nu}$ are rewritten by expanding the Ricci tensor as $R_{\mu\nu}=\overline{R}_{\mu\nu}+R^{(1)}_{\mu\nu}+R^{(2)}_{\mu\nu}+\cdots$ where $\overline{R}_{\mu\nu}$ is the Ricci tensor computed using $\overline{g}_{\mu\nu}$ alone, and $R^{(n)}_{\mu\nu}$ are the correction terms to $R_{\mu\nu}$ and are of the $n$-th order in $h_{\mu\nu}$.
Then, $R^{(1)}_{\mu\nu}$ and $R^{(2)}_{\mu\nu}$ are explicitly given as follows:
\begin{align}
    R^{(1)}_{\mu\nu}
    =&
    \frac{1}{2}
    \left[
    \codev^{\alpha}\codev_{\mu}h_{\nu\alpha}
    +\codev^{\alpha}\codev_{\nu}h_{\mu\alpha}
    -\codev^{\alpha}\codev_{\alpha}h_{\mu\nu}
    -\codev_{\mu}\codev_{\nu}h
    \right],
    \\
    R^{(2)}_{\mu\nu}
    =&
    \frac{1}{2}\overline{g}^{\rho\sigma}\overline{g}^{\alpha\beta}
    \left[
    \frac{1}{2}\codev_{\mu}h_{\rho\alpha}\codev_{\nu}h_{\sigma\beta}
    +
    (\codev_{\rho}h_{\nu\alpha})(\codev_{\sigma}h_{\mu\beta}-\codev_{\beta}h_{\mu\sigma})
    \right.\notag
    \\
    &+h_{\rho\alpha}
    (\codev_{\nu}\codev_{\mu}h_{\sigma\beta}
    +\codev_{\beta}\codev_{\sigma}h_{\mu\nu}
    -\codev_{\beta}\codev_{\nu}h_{\mu\sigma}
    -\codev_{\beta}\codev_{\mu}h_{\nu\sigma}
    )\notag
    \\
    &\left.
    +(\frac{1}{2}\codev_{\alpha}h_{\rho\sigma}-\codev_{\rho}h_{\alpha\sigma})
    (\codev_{\nu}h_{\mu\beta}+\codev_{\mu}h_{\nu\beta}-\codev_{\beta}h_{\mu\nu})
    \right],
\end{align}
where $\codev_{\mu}$ is a covariant derivative with respect to the background metric $\overline{g}_{\mu\nu}$ \citep{Maggiore2018apr}.
Up to quadratic order in $h_{\mu\nu}$, we have the Einstein equations for $\overline{R}_{\mu\nu}$:
\begin{align}
    \overline{R}_{\mu\nu}-\frac{1}{2}\overline{g}_{\mu\nu}\overline{R}
    =&
    8\pi G\left(\overline{T}_{\mu\nu}+t_{\mu\nu}\right),
\end{align}
where $\overline{T}_{\mu\nu}$ is the energy-momentum tensor contributed by matter components, and it varies slowly with time and space, and $t_{\mu\nu}$ is an effective energy-momentum tensor of GWs.
In our case, the derivative of background gravitational potential is small compared to the derivative of GWs due to $L\gg\lambda$.
Under this assumption and by ignoring the derivative of the background potential, the explicit expression of $t_{\mu\nu}$ up to relevant order is given as \citep{Isaacson1968feba,Misner.Thorne.ea1973,Landau.Lifshitz1975jan}:
\begin{align}
    t_{\mu\nu}
    =&
    -\frac{1}{8\pi G} \Braket{
    R^{(2)}_{\mu\nu}-\frac{1}{2}\overline{g}_{\mu\nu}R^{(2)}
    }_{t,x}\notag
    \\
    =&\frac{1}{32\pi G}
    \Braket{
    \overline{g}^{\alpha\rho}\overline{g}^{\beta\sigma}
    \partial_{\mu}h_{\alpha\beta}\partial_{\nu}h_{\rho\sigma}
    -
    \frac{1}{2}
    \overline{g}_{\mu\nu}
     \overline{g}^{\lambda\eta} \overline{g}^{\alpha\rho}\overline{g}^{\beta\sigma}
    \partial_{\lambda}h_{\alpha\beta}\partial_{\eta}h_{\rho\sigma}
    }_{t,x}.
\end{align}
Note that $\braket{\cdots}_{t,x}$ is a space-time average whose integral region is greater than the typical wavelength of GWs and much smaller than the typical scale over which the background metric varies. 
With this definition, it is possible to assign a gauge invariant local energy of GWs. 
Now, we introduce the polarization tensor $e_{\mu\nu}$ such that $h_{\mu\nu}=\phi e_{\mu\nu}$ ($e_{\mu\nu}e^{\mu\nu}=2, e^{\mu}_{\mu}=0$) and by setting $e_{\mu\nu}$ to a constant \citep{Misner.Thorne.ea1973,Peters1974apr}, we obtain the following:
\begin{align}
     t_{\mu\nu}=
    \frac{1}{16\pi G}
    \Braket{
    \partial_{\mu}\phi\partial_{\nu}\phi
    -\frac{1}{2}\overline{g}_{\mu\nu}
    \partial_{\lambda}\phi\partial^{\lambda}\phi
    }_{t,x}.
\end{align}
Using this notation, the total energy of GWs in volume $V$ averaged out over a certain period of time $T$, denoted as $\braket{\cdots}_{t}=(1/T)\int_{t}^{t+T} dt'(\cdots)$, is given by 
\begin{align}
    \label{ap: energy}
    E=&\int\braket{ t^{00}}_{t}dV\notag
    \\
    =& \frac{1}{16\pi G}\int_{t}^{t+T}\frac{dt'}{T}\int dV
    \left(\frac{1}{2}(\nabla \phi)^{2}+\frac{1}{2}\dot{\phi}^{2}-2\Phi\dot{\phi}^{2}\right).
\end{align}
By combining the conservation of energy $\partial_{\mu}t^{\mu\nu}=0$, we obtain the following:
\begin{align}
    \partial_{0}E
    =&-\int\partial_{i}\braket{t^{0i}}_{t}dV\notag
    \\
    =&-\frac{1}{16\pi G}\int_{t}^{t+T}\frac{dt'}{T}\int_{S} dS n_{i}\cdot \left(
    -\dot{\phi}\partial_{i}\phi
    \right).
\end{align}
Note that the space-time average $\braket{\cdots}_{t,x}$ is removed when $(1/T)\int_{t}^{t+T}dt'\int dV$ is taken.
This expression is the same as the one derived in section~\ref{sec: consistency} using the wave equation (\ref{eq: wave equation}).
Thus, the conserved quantity associated with Eq.~(\ref{eq: wave equation}) is properly considered as the energy of GWs.

Note that, only one degree of freedom associated with the polarization of GWs is considered in this discussion. 
When accounting for two polarization components ($h_{\mu\nu}=\phi_{\times}e^{\times}_{\mu\nu}+\phi^{+}e_{\mu\nu}^{+}$) and assuming that the polarization tensors $e^{\times}_{\mu\nu}$ and $e^{+}_{\mu\nu}$ are independent, 
the total energy of GWs is simply given by the sum of the energy of the 
$\times$ mode $E^{\times}$ and the $+$ mode $E^{+}$, i.e., $E=E^{\times}+E^{+}$.

\bibliography{Consistency_Relation}
\bibliographystyle{JHEP}

\end{document}